\documentclass[letter]{aa}
\usepackage[varg]{txfonts}
\usepackage{graphicx}
\usepackage{t1enc}
\usepackage{epsfig}
\usepackage{rotating}
\usepackage[authoryear]{natbib}
\bibpunct{(}{)}{;}{a}{}{,}

\newcommand{\Tex}   {$T_\mathrm{ex}$}
\newcommand{\Trot}  {$T_\mathrm{rot}$}

\newcommand{\kms}   {km~s$^{-1}$}
\newcommand{\cmt}   {cm$^{-3}$}
\newcommand{\cmd}   {cm$^{-2}$}
\newcommand{\jpb}   {$\mathrm{Jy~beam^{-1}}$} 
\newcommand{\lo}    {$L_{\sun}$}

\newcommand{\nh}    {NH$_3$}
\newcommand{\nth}   {N$_2$H$^+$}
\newcommand{\nhtd}  {NH$_2$D}

\newcommand{\et}    {et al.}
\newcommand{\eg}    {e.\,g.,}
\newcommand{\ie}     {i.\,e.,}
\newcommand{\uchii} {UC\ion{H}{ii}}

\newcommand{\velo} {$v$}
\newcommand{\dfrac}	{$D_{\mathrm{frac}}$}

\newcommand{\hii}	{\ion{H}{ii}}

\newcommand{\phnp}   {\phantom{0.}}

\newcommand{\phn}   {\phantom{0}}

\newcommand{\php}   {\phantom{.}}

\begin{document}

	\title{The \nhtd/\nh\ ratio toward pre-protostellar cores around\\ the UC\hii\ region in IRAS\,20293+3952\thanks{Based on observations carried out with the IRAM Plateau de Bure Interferometer. IRAM is supported by INSU/CNRS (France), MPG (Germany), and IGN (Spain).}}
	
	\author{G. Busquet\inst{1} 
       	 \and
	 Aina Palau\inst{2}
	 \and
	 R. Estalella\inst{1}
	 \and
	 J.~M. Girart\inst{2}
	 \and
	 \'A. S\'anchez-Monge\inst{1}
	 \and
	 S. Viti\inst{3}
	 \and
	 P.~T.~P. Ho\inst{4,5}
	 \and
	 Q. Zhang\inst{4}}

  	\offprints{Gemma Busquet,\\ \email{gbusquet@am.ub.es}}

	\institute{Departament d'Astronomia i Meteorologia (IEEC-UB), Institut de 
	Ci\`encies del Cosmos, Universitat de Barcelona, Mart\'{\i} i Franqu\`es 1, E-08028 Barcelona, Spain
       \and
        Institut de Ci\`encies de l'Espai (CSIC-IEEC), Campus UAB,
         Facultat de Ci\`encies, Torre C-5 parell, E-08193 Bellaterra, Spain
         \and
	Department of Physics and Astronomy, University College London, Grower Street,
	 London WC1E 6BT, UK
	\and
	 Harvard-Smithsonian Center for Astrophysics, Cambridge, MA, 02138, USA
			 \and
	 Academia Sinica Institute of Astronomy and Astrophysics, Taipei, Taiwan
	       }
 	 \date{Received / Accepted}

	  \authorrunning{G. Busquet \et}
 	 \titlerunning{Deuterated ammonia in IRAS\,20293+3952}

\abstract{The deuterium fractionation, \dfrac, has been proposed as an evolutionary indicator in
pre-protostellar and protostellar cores of low-mass star-forming regions.}{We investigate \dfrac, with high
angular resolution, in the cluster environment surrounding the UC\hii\ region
IRAS\,20293+3952.}{We performed high angular resolution observations  with the IRAM Plateau de Bure
Interferometer (PdBI) of the ortho-\nhtd\,$1_{11}$--$1_{01}$ line at 85.926~GHz and compared them with
previously reported VLA \nh\ data.}{We detected strong \nhtd\ emission toward the pre-protostellar cores
identified in \nh\ and dust emission, all located in the vicinity of the \uchii\ region
IRAS\,20293+3952. We found high values of  \dfrac$\simeq$0.1--0.8 in all the pre-protostellar cores and
low values, \dfrac$<0.1$, associated with young stellar objects.}{The high values of \dfrac\ in
pre-protostellar cores could be indicative of evolution, although outflow interactions and UV radiation
could also play a role.}


\maketitle

\section{Introduction}

Characterizing the different evolutionary stages before and after the
formation of a star is crucial for understanding the process of
star formation itself. Dense cores, where stars are born, are mainly
studied through molecular emission of dense gas tracers, and N-bearing
molecules are widely used because they do not freeze out onto dust
grains until very high densities are reached ($\sim10^6$~\cmt;
\citealt{flower2006}).  Several column density ratios of
molecules tracing dense cores have been proposed as good chemical
clocks, such as \nh/\nth\ or CN/\nth\
\citep{hotzel2004,fuente2005}. Among these, the ratio of a
deuterated species over its counterpart containing H, \ie\ the
deuterium fractionation \dfrac, has been found to be a good tracer of the evolutionary stage of dense cores. Both observations and models strongly suggest that in cold ($T<20$~K) and dense cores ($n\simeq10^6$~\cmt) C-bearing molecules are expected to deplete onto dust grains, leading to an enhancement of the deuterium fractionation \citep[\eg][]{roberts2000,bacmann2003,pillai2007} because the H$_2$D$^+$ ion, the progenitor of most of the deuterated species, including \nhtd, is not destroyed by CO. 

In particular, in low-mass star-forming regions, \dfrac\ is found to
increase until the onset of star formation and to decrease afterwards
\citep{crapsi2005,emprechtinger2009}. In the high-mass regime,
\citet{chen2010} performed single-dish observations toward massive
dense cores and find a decreasing trend
of \dfrac\ with the evolutionary stage during the protostellar
phase. However, since massive star formation takes place in cluster environments, to study \dfrac\ one needs to carry out observations with high angular resolution in
order to separate the emission from each individual core. So far, interferometric observations of deuterated species in massive star-forming regions have only been reported by \citet{fontani2008} and \citet{sandell2010}, but only the work of \citet{fontani2008} evaluates \dfrac\ in the region, where they find two dense cores with \dfrac$\simeq0.1$. 

\begin{figure}[t]
\begin{center}
\begin{tabular}{c}
    \epsfig{file=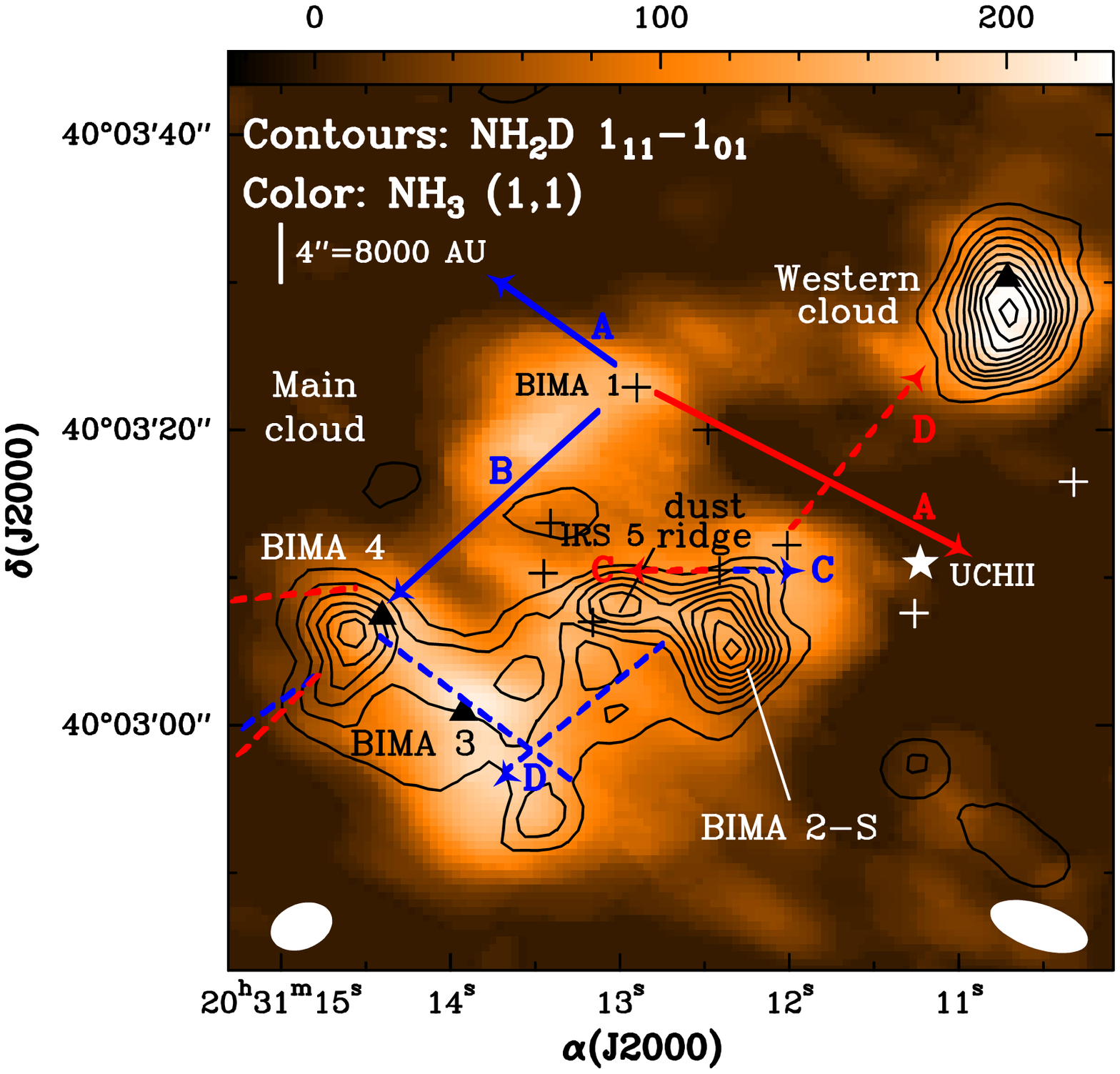,scale=0.48}\\
    \end{tabular}
     \caption{Contours: \nhtd\  zero-order moment. Contours start at 6~\%, increasing in steps of 10~\% of the peak intensity, 0.58~\jpb~\kms. Color scale: \nh\,(1,1) zero-order moment \citep{palau2007}. The \nhtd\ synthesized beam, 4$\farcs32\times3\farcs05$ (P.~A.$=111\degr$), and the \nh\ synthesized beam are shown in the bottom left and bottom right corners, respectively. White star: position of the \uchii\ region (IRAS\,20293+3952). Plus signs: YSOs embedded in the main cloud. Black triangles: \nh\ column density peaks in the starless cores BIMA~3, BIMA~4, and the western cloud. Main cores of \nhtd\ are labeled as western cloud, dust ridge, BIMA~2-S, and BIMA~4. The blue and red continuum/dashed arrows mark the direction of powerful/less powerful molecular outflows, respectively \citep{beuther2004a}. Outflow B is deflected by BIMA~4 \citep{palau2007} and the dashed lines indicate the high-velocity gas around BIMA~4.}
\label{fnh2dmom0}
\end{center}
\end{figure}

In this letter we present high angular resolution observations
of the ortho-\nhtd~$1_{11}$--$1_{01} $ line at 85.926~GHz carried out
with the Plateau de Bure Interferometer toward the high-mass
star-forming region IRAS\,20293+3952, located at 2~kpc of
distance \citep{beuther2004b} and with a luminosity of 6300~\lo. The region is associated with an \uchii\ region at the border of a cloud of dense
gas mapped in \nh\ with high angular resolution
\citep{palau2007}, and it harbors a rich variety of young stellar objects (YSOs) and dense cores at
different evolutionary stages, being thus an excellent target to study \dfrac. While the northern side of the main
cloud of the region contains several YSOs driving at least four
molecular outflows \citep{beuther2004a,palau2007}, the southern side
is mainly populated with starless core candidates (BIMA~3 and BIMA~4, two faint
millimeter condensations). In addition, there is a
small cloud to the northwest, the western cloud, which seems to harbor
a very compact starless core (see Fig.~\ref{fnh2dmom0} for an overview
of the region). The current \nhtd\ observations allowed us to estimate
for the first time \dfrac\ from \nhtd/\nh with high angular resolution toward a
massive star-forming region containing dense cores and YSOs at
different evolutionary stages.

\section{Observations}

The IRAM Plateau de Bure Interferometer (PdBI) was used to observe the  \nhtd\,1$_{11}-1_{01}$ molecular transition at 85.926~GHz toward IRAS\,20293$+$3952. The observations were carried out
during 2008 June 16 and December 4, with the array in the D (4 antennas) and C (6 antennas) configurations,
respectively, providing projected baselines ranging from 17.5~m to 175~m. The phase center was 
$\alpha(\mathrm J2000)=20^{\mathrm h}31^{\mathrm m}12\fs7$, $\delta(\mathrm
J2000)=+40\degr03\arcmin13\farcs4$. The typical system temperatures for the receivers at 3~mm were
$\sim$150--300~K during June 16, and $\sim$100--150~K during December 4. Atmospheric phase
correction was applied. 

The 3~mm receiver was tuned at 86.75433~GHz in the lower sideband. The spectral
setup included other molecular transitions, which will be presented in a subsequent paper (Busquet et al., in prep). For the ortho-\nhtd, we used a
correlator unit of 20~MHz of bandwidth and 512 spectral channels, which provides a spectral resolution of
$\sim0.039$~MHz ($\sim0.14$~\kms). The FWHM of the primary beam at 3~mm was $\sim56''$. Bandpass calibration was performed by observing the quasars 3C454.3 on June 16 and 3C273 on December 4.
Amplitude and phase calibrations were achieved by monitoring MWC\,349 and 2005$+$403 for both days. The phase rms was $\sim20\degr$. The absolute flux density scale was determined from MWC\,349, with an
estimated uncertainty $\sim10$\%. The data were calibrated with the program CLIC and imaged with MAPPING, and
both are part of the GILDAS software package. Imaging was performed with natural weighting, obtaining a synthesized beam of $4\farcs32\times3\farcs05$ with P.~A.$=111\degr$, and rms of 20.7~m\jpb\ per channel.

\begin{figure*}[t]
\begin{center}
\begin{tabular}{cc}
    \epsfig{file=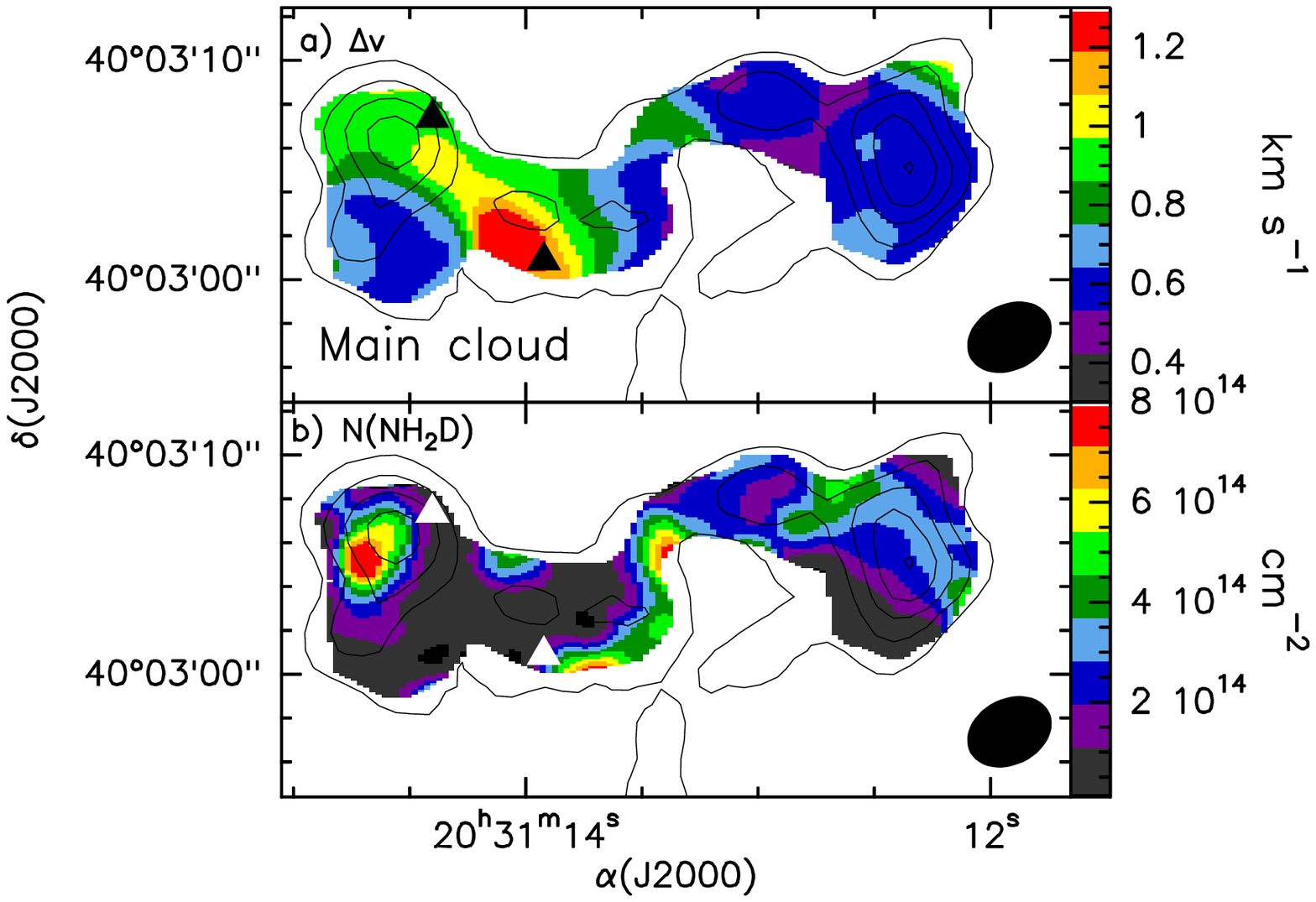,scale=0.5}&
    \epsfig{file=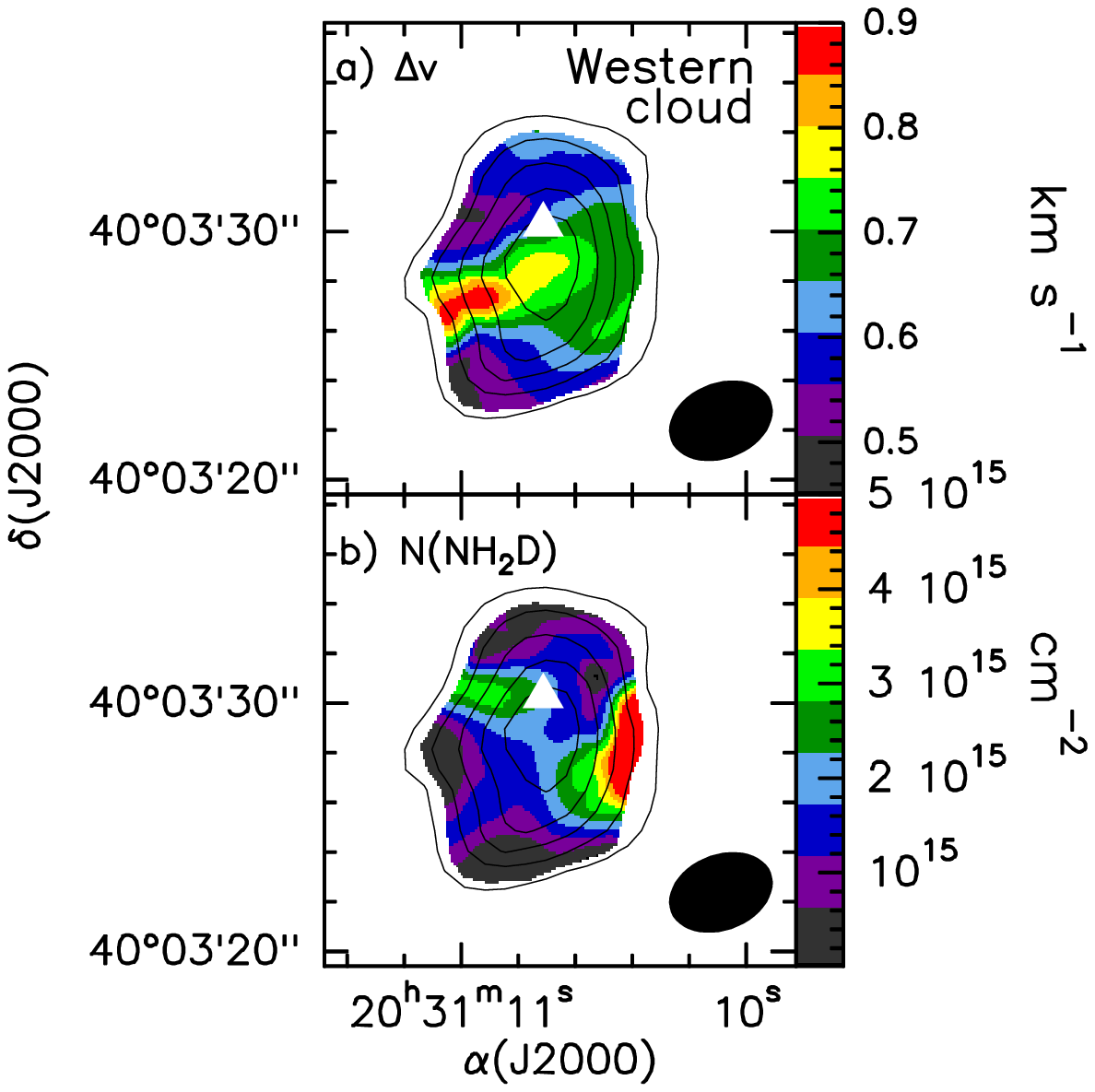,scale=0.5}\\
    \end{tabular}
     \caption{Color-scale maps of linewidth and $N$(\nhtd) overlaid on the \nhtd\,$1_{11}$-$1_{01}$ integrated emission (contours) for the main cloud (left panels) and the western cloud (right panels). Triangles mark the \nh\ column density peaks of BIMA~3, BIMA~4 (left), and the western cloud (right). In all panels the synthesized beam is shown in the bottom right corner.}
\label{fnh2dpar}
\end{center}
\end{figure*}

\begin{figure*}[!ht]
\begin{center}
\begin{tabular}{c}
    \epsfig{file=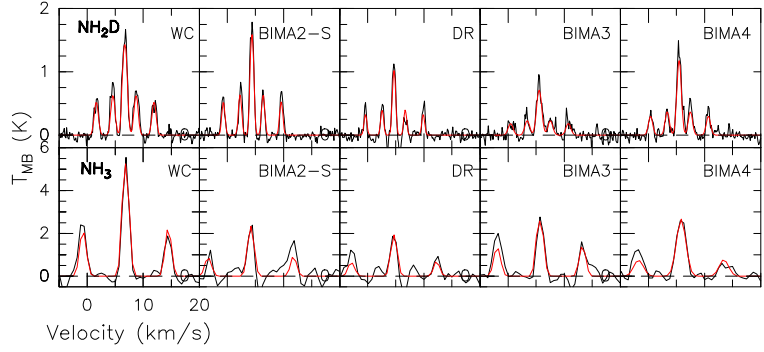,scale=0.71}\\
    \end{tabular}
     \caption{Spectra (black line) toward 5 positions of the IRAS\,20293+3952 for \nhtd\,$1_{11}$-$1_{01}$ (top) and \nh\,(1,1) (bottom). The 5 positions, which are labeled in the top right corner of each panel, are, from left to right WC (western cloud), BIMA~2-S, DR (dust ridge core), BIMA~3, and BIMA~4. The red lines show the fit to the hyperfine structure obtained as illustrated in Sect.~3.}
\label{fdfrac}
\end{center}
\end{figure*}

\section{Results}

In Fig.\ref{fnh2dmom0} we present the \nhtd\,$1_{11}$-$1_{01}$ zero-order map integrated for the 6 hyperfine transitions (see \citealt{olberg1985} for a description of the transition) overlaid onto the \nh\,(1,1) emission from \citet{palau2007}. The integrated \nhtd\  emission toward the western cloud  presents a compact morphology with a deconvolved size of $6\farcs5\times3\farcs6$ ($\sim0.06$~pc). We also found \nhtd\ emission associated with the southern side of the main cloud, covering a spatial extension in the east-west direction of $\sim33''$ (0.33~pc). The emission consists of three main cores, one clearly associated with BIMA~4,  another \nhtd\ core coincident with the dust ridge, and a third core located $\sim7''$ south of BIMA~2 (hereafter BIMA~2-S). In addition, we detected faint \nhtd\ emission associated with BIMA~3. It is worth noting that \nhtd\ is not (or marginally) detected on the northern side of the main cloud. In this letter we follow the nomenclature for the YSOs and starless core candidates of \citet{palau2007}.

We studied the deuterium fractionation of the region by first computing the column density map of \nhtd\ and then the $N$(\nhtd)/$N$(\nh) column density ratio. 
For this, we extracted the spectra for positions on
a grid of $2''\times2''$, and adopted the hyperfine frequencies listed in \citet{tine2000}. The column densities are only reported for
spectra with a peak intensity greater than 5~$\sigma$, and with all
the hyperfine components detected.

Assuming that all levels are populated according to the same excitation temperature, \Tex, the column density of the asymmetric top molecule  \nhtd\ is given by

\begin{small}
\begin{eqnarray}
\Big[\frac{N(\mathrm{NH_{2}D})}{\mathrm{cm^{-2}}}\Big]&=&1.94\times10^{3}\Big[\frac{\nu}{\mathrm{GHz}}\Big]^{2}\Big[\frac{A_{ul}}{\mathrm{s}^{-1}}\Big]^{-1}\frac{Q(T_{\mathrm{ex}})}{g_{u}}\exp
\Big(\Big[\frac{E_{u}}{\mathrm{K}}\Big]\Big[\frac{T_{\mathrm{ex}}}{\mathrm{K}}\Big]^{-1}\Big)
\nonumber\\
&& {}\times~J_{\nu}(T_{\mathrm{ex}})\tau_{m}\Big[\frac{\Delta~v}{\mathrm{km~s^{-1}}}\Big],
 \label{eqcoldens}
 \end{eqnarray}\end{small}where $\Delta$\velo\ is the linewidth, $\tau_{m}$ the optical depth of the main line ($1_{11}$--$1_{01}$, $F$=2--2), $\nu$ the frequency of the transition ($\nu=85.926$~GHz), $E_{u}$ the energy of the upper level ($E_{u}=20.68$~K), $A_{ul}=5.8637\times10^{-6}$~s$^{-1}$ is the Einstein coefficient, and $J_{\nu}(T_{\mathrm{ex}})=(h\nu/k)/(e^{h\nu/kT_{\mathrm{ex}}}-1)$, where $h$ and $k$ are the Planck and Boltzmann constants, respectively . From the Cologne Database for Molecular Spectroscopy (CDMS; \citealt{muller2001}), the upper level degeneracy is $g_{u}=15$, and the partition function $Q$(\Tex) is estimated as $Q$(\Tex)$=\alpha+\beta$\Tex$^{3/2}$, where $\alpha=3.899$ and $\beta=0.751$ are the best-fit parameters from a fit to the partition function at the different temperatures given in CDMS. The standard ortho/para ratio of 3 \citep[\eg][]{walmsley1987,tine2000} is already included in the partition function.

\begin{table*}[t]{\center
\caption{Summary of the main physical properties of selected cores in IRAS\,20293+3952
}
\begin{small}
\begin{tabular}{lcccccccccc}
\hline\hline
& $\alpha$(J2000)& $\delta$(J2000)  &\Tex$^\mathrm{a}$ & &\Trot$^\mathrm{c}$  & $N$(\nhtd)$^\mathrm{d}$ &$N$(\nh)$^{\mathrm{d}}$ &  &\nh/ &Evol.   \\
Core &(h~m~s) &($\degr$ $'$ $''$) &(K) & $\tau_{m}^{\mathrm{b}}$ &(K)  & ($\times10^{14}$~\cmd) &($\times10^{14}$~\cmd) &$D_{\mathrm{frac}}^{\mathrm{e}}$ &\nth$^{\mathrm{c}}$  &stage$^{\mathrm{f}}$  \\
\hline
Western cloud  &20:31:10.70 &40:03:28.2   &4.8$\pm0.2$  &1.5  &15$\pm3$   &\php10$\pm\phnp2$ &\php14$\pm\phnp2$ &0.8$\pm0.2$ &300 &PPC   \\  
BIMA~2-S &20:31:12.34 &40:03:05.7           &5.4$\pm0.3$  &1.1   &19$\pm3$   &2.8$\pm0.6$       &6.7$\pm0.8$&0.4$\pm0.1$   &\phn90  &PPC \\
Dust ridge  &20:31:12.98 &40:03:08.5          &4.6$\pm0.3$  &1.0   &16$\pm3$  &2.8$\pm0.7$       &4.9$\pm0.9$&0.6$\pm0.2$  &\phn90  &PPC\\
BIMA~3 &20:31:13.92 &40:03:00.9              &5.4$\pm1.0$   &0.4   &14$\pm2$  &1.9$\pm0.5$        &\php19$\pm\phnp3$ &0.10$\pm0.04$ &300 &PPC \\
BIMA~4    &20:31:14.56 &40:03:06.8         &7.5$\pm1.6$     &0.3   & 14$\pm2$  &1.2$\pm0.4$       &7.4$\pm0.8$ &0.16$\pm0.05$  &200  &PPC\\
BIMA~1 &20:31:12.77 &40:03:22.6 &\ldots\  &\ldots\ &24$\pm3$  &$<1.1^\mathrm{g}$ &19$\pm3$ &$<0.06$ &\phn50  &O, no IR\\
IRS~5 &20:31:13.41 &40:03:13.7 &\ldots\   &\ldots\  &24$\pm3$ &$<1.5^\mathrm{g}$ &11$\pm2$ &$<0.1$ &\phn50  &IR\\
\hline
\end{tabular}
\end{small}
\begin{list}{}{}
\textbf{Notes.} $^\mathrm{(a)}$ \Tex\ derived from the output parameters of the \nhtd\,$1_{11}$-$1_{01}$ hyperfine fit. $^\mathrm{(b)}$ Derived from the fits to the hyperfine structure of \nhtd\,$1_{11}$-$1_{01}$. $^\mathrm{(c)}$ Obtained from \citet{palau2007}. $^\mathrm{(d)}$ Column densities derived using the same $uv$ range and convolving the images to a circular beam of $7''$. $^\mathrm{(e)}$ $D_{\mathrm{frac}}=N$(\nhtd)/$N$(\nh). $^\mathrm{(f)}$ PPC: pre-protostellar core; O: molecular outflow.
$^\mathrm{(g)}$ 3$\sigma$ upper limit estimated adopting $\Delta\,v$ from \nh\ and \Tex$=5$~K.
\end{list}
\label{tdfrac}
}
\end{table*}

Figure~\ref{fnh2dpar} shows a map of the linewidth ($\Delta$\velo) derived from the hyperfine fit and the total column density $N$(\nhtd) toward the southern side of the main cloud (\ie\ dust ridge core, BIMA~2-S, BIMA~3, and BIMA~4, hereafter referred to as the main cloud; see left panels) and the western cloud (right panels). The properties of the western cloud and the main cloud are somewhat different. The total optical depth ($\tau_{1_{11}-1_{01}}=2\tau_{m}$) in the western cloud is in the range $\sim4$--15, with a typical uncertainty of $\sim1.1$, whereas the total optical depth is lower toward the main cloud, between 0.5 and 4, with a typical uncertainty of $\sim0.5$.  Concerning the linewidth in these clouds (see Fig.~\ref{fnh2dpar}\,a), we find line broadening  ($\Delta$\velo$\simeq0.75$--0.9~\kms) associated with the western cloud. Toward BIMA~3 and BIMA~4, we also find high values for the linewidth,  around  $\sim0.8$--1.2~\kms, while the dust ridge core and BIMA~2-S appear to be more quiescent, with linewidths of $\sim$0.5--0.7~\kms.  Since it has been suggested that one of the outflows of the region, outflow~B, is interacting with  BIMA~4 (Palau et al. 2007), this could produce the line broadening seen in \nhtd\ in BIMA~3 and BIMA~4. Additionally, the line broadening found in the western cloud is spatially coincident (in projection) with a high-velocity feature of outflow~D  \citep{beuther2004a}. Thus, the broad lines seen toward the western cloud, BIMA~3, and BIMA~4, suggest that the deuterated gas is being perturbed by the passage of the outflows (see Fig.~\ref{fnh2dmom0}). 

The \nhtd\ column density, corrected for the primary beam response, also presents significant differences between the western cloud and the main cloud. While the column density is, on average, $N$(\nhtd)$\simeq25\times10^{14}$~\cmd\ in the western cloud, in the main cloud the column density is slightly lower, in the range $1$--8$\times10^{14}$~\cmd, reaching its maximum value of $8\times10^{14}$~\cmd\ close to the peak position of BIMA~4 (see Fig.~\ref{fnh2dpar}\,b). The uncertainty of the \nhtd\ column density is $\sim$25--35~\%. 

In order to properly estimate the deuterium fractionation, defined as $D_{\mathrm{frac}}=N$(\nhtd)$/N$(\nh), we made the \nh\ VLA images using the same $uv$ range as the PdBI data (5--50~$k\lambda$), estimating the column densities for both \nh\ and \nhtd\ for the same angular scales. Finally, we convolved the \nhtd\ and \nh\ emission to a circular beam of $7''$ (the major axis of the \nh\ beam). In Fig.~\ref{fdfrac} we present the spectra obtained at the \nhtd\ emission peak of each condensation (\ie\ western cloud, BIMA~2-S, dust ridge core, BIMA~3, and BIMA~4) together with the hyperfine fit obtained toward these positions.  In Table~\ref{tdfrac} we list the excitation temperature, \Tex, the rotational temperature, \Trot, the \nhtd\ and \nh\ column densities, $D_{\mathrm{frac}}$, and the \nh/\nth\ ratio for each core and a few YSOs. Toward the YSOs BIMA~1 and IRS~5, we report on upper limits, with \dfrac$<0.1$, and we cannot draw any conclusion for the behavior of \dfrac\ in the protostellar phase. More interestingly, in the western cloud \dfrac\ is $\sim0.8$, which is the highest value of \dfrac\ in the region and among the highest reported in the literature \citep[\eg][]{crapsi2007,pillai2007,fontani2008}. In the main cloud, \dfrac\ presents significant variations among the different cores, with \dfrac\ decreasing from the northwest (\dfrac$\simeq$0.5 in BIMA~2-S and the dust ridge core) to the southeast (\dfrac$\simeq$0.1 in BIMA~3 and BIMA~4). This suggests a chemical differentiation along the main cloud.

\section{Discussion and summary}

Our high angular resolution study of the \nhtd\ toward the massive star-forming region IRAS\,20293+3952 reveals strong \nhtd\ emission toward starless cores, whereas \nhtd\ is not (or marginally) detected in cores containing YSOs, which suggests that the production of \nhtd\ is more effective in the pre-protostellar phase than in the protostellar phase. 
\citet{palau2007} notice that the starless cores in this region seem to be predominant on the southern side of the main cloud and in the western cloud, while the northern side of the main cloud harbors all the YSOs known in the region, suggestive of the dense gas in the main cloud being progressively more evolved as it moves from south to north. In addition, chemical differentiation among pre-protostellar and protostellar cores was also found by \citet{palau2007} using the \nh/\nth\ ratio, which 
was high for pre-protstellar cores and low in protostellar cores (see col.~9 of Table~\ref{tdfrac}). Thus, for this region, the behavior of \dfrac, measured from \nhtd/\nh, is similar to the behavior of \nh/\nth\ ratio, suggesting that both ratios can be used to distinguish between pre-protostellar and protostellar cores and that both ratios could be related with the evolutionary stage of the dense gas. 

A possible interpretation of the differences in \dfrac\ seen in the pre-protostellar cores of region could be that they are in different evolutionary stages. According to the study of \citet{crapsi2005}, there is an increasing trend for \dfrac\ as the starless core approaches the onset of gravitational collapse (from 0.03--0.1 in the youngest cores to 0.1--0.4 toward the most evolved cores). This would indicate that the western cloud is the most evolved pre-protostellar core and that BIMA~3 and BIMA~4 are less evolved. However, in regions of massive star formation, typically associated with clustered environments, other factors, like temperature, UV radiation, and/or molecular outflows, can play important roles in altering the chemistry, and then it is not clear whether this trend is related to the evolutionary stage of pre-protostellar cores. 

A comparison of \dfrac\ between the western cloud and BIMA~3/BIMA~4, all of them having similar temperatures but very different values of \dfrac\ (see Table~\ref{tdfrac}), indicates that temperature is not an important factor in determining \dfrac\ for temperatures around 15~K. \citet{palau2007} find evidence that UV radiation from the \uchii\ region affects the chemistry at the western edge of the main cloud facing the \uchii\ region, so it could affect BIMA~2-S and the western cloud. In particular, the presence of a cavity between the \uchii\ region and the western cloud suggests that this cloud could be photo-illuminated by the \uchii\ region. However, while the western cloud and BIMA~2-S could be locally affected by the UV radiation, this is not the case for BIMA~3 and BIMA~4, for which the high visual extinction in the main cloud prevents UV photons from penetrating. Finally, \citet{fontani2009} print out that shocks in outflows could modify \dfrac. This could be the case for the western cloud, BIMA~3, and for BIMA~4. While the interaction of an outflow with the western cloud is not evident, \citet{palau2007} have already proposed that the powerful outflow~B is interacting with BIMA~4, producing the deflection of the outflow and thus the ejection of high-velocity material in different directions. This could affect BIMA~4 and BIMA~3 because high-velocity SiO is seen close to these cores (see Fig.~\ref{fnh2dmom0} and \citealt{beuther2004a}). This interpretation is reinforced by the line broadening of  \nhtd\ observed towards BIMA~3 and BIMA~4 (see Fig.~\ref{fnh2dpar}).  In conclusion, while \dfrac\ in IRAS\,20293+3952 may be locally affected by the interaction of outflows and UV radiation, \dfrac\ is lower toward YSOs than toward pre-protostellar cores, with a  possible evolutionary trend in the pre-protostellar phase, which deserves further study.

\begin{acknowledgements} 

We thank the anonymous referee for his/her useful comments and suggestions. We acknowledge the IRAM staff for their support 
during the data reduction. G.~B., A.~P., R.~E.,
J.~M.~G., and \'A.~S.-M. are supported by the Spanish MEC grant AYA2005-08523-C03 and the MICINN grant AYA2008-06189-C03 (co-funded with FEDER funds). A.~P. is supported by a JAE-Doc CSIC fellowship co-funded with the 
European Social Fund. 

\end{acknowledgements}

\end{document}